\documentclass[aps,twocolumn,floatfix]{revtex4}
\usepackage{epsfig}
\newcommand{\be}{\begin{equation}}
\newcommand{\ee}{\end{equation}}

\begin{document}
\title{Heat Conduction in Low Dimensions: From Fermi-Pasta-Ulam Chains 
        to Single-Walled Nanotubes}
\author{Peter Grassberger and Lei Yang}
\affiliation{John-von-Neumann Institute for Computing, Forschungszentrum J\"ulich,
D-52425 J\"ulich, Germany}

\date{\today}

\begin{abstract}
Heat conduction in 1-dimensional anharmonic systems is anomalous in 
the sense that the conductivity $\kappa$ scales with a positive power 
of the system size, $\kappa\sim L^\alpha$. In two dimensions, previous 
simulations and theoretical arguments gave a logarithmic divergence. 
For rectangular systems of size $L_\|\times L_\perp$ there should be 
a cross-over from the 2-d to the 1-d behaviour as the aspect ratio 
$r = L_\|/ L_\perp$ increases from $r=1$ to $r\gg 1$. When taking 
periodic boundary conditions in the transverse direction, this should 
be of direct relevance for the heat conduction in single-walled carbon 
nanotubes. In particular, one expects that $k$ nanotubes of diameter 
$R$ should conduct heat better than a single
nanotube of the same length and of radius $kR$. We study this cross-over 
numerically by simulating the Fermi-Pasta-Ulam model. Apart from giving 
a precise estimate of the exponent $\alpha$, our most intriguing results
are that the divergence does not seem to be logarithmic in $d=2$ but also
power-like, and that the cross-over does not happen at a fixed aspect 
ratio. Instead, it happens at $r=r^*$ with $r^*\to \infty$ for $L\to\infty$.
\end{abstract}

\maketitle

After years of intense studies it is now clear that heat conduction in
typical anharmonic systems is anomalous in low dimensions \cite{lepri-review}.
In particular, in 1-d systems like the Fermi-Pasta-Ulam (FPU) $\beta$-model
\cite{beta}, the diatomic Toda chain \cite{hatano}, or in 1-d hard-particle 
gases with alternating masses \cite{hatano,dhar,grass-lei,casati-prozen} the 
heat conductivity $\kappa$, defined by $J(x) = \kappa \nabla T(x)$, scales
as 
\begin{equation}
 \kappa \sim L^\alpha                       \label{kappa}
\end{equation}
where $L$ is the size of the system. Notice that
we have assumed here that $T(x)$ exists, i.e. that local thermal equilibrium
(LTE) is established in the limit $L\to\infty$, which is non-trivial 
\cite{dhar2} but seems to be true for these systems.

The main reason for this anomalous behaviour is the fact that the free path
length of phonons with long wave lengths $\lambda$ diverges for $\lambda 
\to \infty$. Thus soft modes propagate nearly ballistically. In higher 
dimensions this is also true, but due to the enlarged phase space for 
other modes it is restricted to such a small region that $\kappa$ remains 
finite. In 1 dimension there are just not enough other modes with which
soft phonons could interact to make $\kappa$ finite. This argument 
requires of course that there is an acoustic phonon branch, i.e. that 
soft (Goldstone) modes exist because translation invariance is not 
broken. Thus this argument does not apply to charge conduction where the 
electrons move in an external potential braking translation invariance.

For 2-d systems, simulations of FPU lattices of rather modest sizes 
\cite{lippi} indicated a logarithmic divergence $\kappa \sim \ln L$. This 
was also supported by mode coupling theory. Indeed, a logarithmic divergence 
should not be too surprising in view of the logarithmic divergence of 
transport coefficients in 2-d hydrodynamics \cite{evans} due to long 
time tails. In \cite{lippi} it was also observed that the cross-over from
the 1-d to the 2-d scaling in rectangular systems happens at surprisingly 
large aspect ratios. Consider a system of size $L_\|\times L_\perp$ where
the length $L_\|$ is parallel to the heat flow. The aspect ratio is 
defined as $r = L_\|/ L_\perp$. For the sizes studied in \cite{lippi}, 
the cross-over happened at $r\approx 10$, but no detailed study was 
made.

The main aim of the present paper is to present detailed simulations 
of the FPU $\beta$-model on much larger lattices than in \cite{lippi},
in order to study in detail the asymptotic behaviour of 2-d systems
and of the cross-over. But before doing so, let us point out that this 
study is not entirely academic. It is of immediate experimental and maybe 
even of technological importance.

The lattices which we shall use have periodic boundary conditions in the 
transverse direction. Therefore, we actually study the heat conduction 
through (single-walled) tubes. By far the most important nanotubes are 
those made of carbon \cite{nanotubes}. Their heat conduction is dominated 
by phonons. They can be made with very few lattice defects, such that 
their conductivity is basically controlled by phonon-phonon interactions 
as in the FPU model. Indeed, this conductivity has been both calculated 
\cite{nano-sim} and measured \cite{nano-exp}, and is found to be very large. 
It was thus suggested that carbon nanotubes should be the ideal material to 
carry away Joule heat in the next generation of integrated circuits which 
will use nanoscale structures and for which cooling will be a major 
problem. Since most molecular dynamics simulations and measurements were
done for fixed tube lengths, no length dependence of $\kappa$ was seen 
in them. The only exception is Ref.\cite{maruyama} where a clear length
dependence of $\kappa$ was seen for the narrowest tubes, but not for 
wider ones. 

In all simulations of carbon nanotubes the realistic Tersoff-Brenner 
potential \cite{brenner} was used. If, as we claim, the size dependence
of $\kappa$ is universal, its main features (and in particular the 
exponent $\alpha$ in Eq.(\ref{kappa})) should be independent of the 
potential. There will be of course unknown scale factors when transferring
our results from FPU systems to carbon tubes, nevertheless we should be able 
to use our results directly for an experimentally accessible system.

In our FPU simulations we used a square lattice with sites indexed by
integer vectors ${\bf n} = (i,j)$ with $1\le i \le N_x$ and $1\le j \le N_y$. 
We use periodic b.c. in $j$. The lattice constant, the particle mass, and 
the $\beta$ parameter are all set to unity. Thus $N_x=L_\|$ and 
$N_y=L_\perp$ and, apart from the heat baths at $i=1$ and $i=N_x$, the 
hamiltonian is 
\be
   H = {1\over 2}\sum_{\bf n} p_{\bf n}^2 + \sum_{<{\bf n, m}>} 
       \left[{1\over 2}(q_{\bf n}-q_{\bf m})^2 + {1\over 4}(q_{\bf n}-q_{\bf m})^4
       \right]\;.
\ee
Notice that identically the same form (with $N_y=1$) can be used also for 
1-d chains. To measure a heat flux we keep the left boundary ($i=1$) at a 
lower temperature $T_h = 10$, and the right boundary ($i=N_x$) at $T_c = 6$.
These are fairly large in order to obtain fast equilibration (similar temperatures
were used in \cite{beta}). These thermostats are implemented as Nos{\'e}-Hoover
\cite{evans} heat baths with response times $\Theta=1$. Thus the equations 
of motion for the particles in the boundaries are modified to
\be
   \ddot{q}_{\bf n} = - {\partial H \over \partial q_{\bf n}} -
         \xi_{\bf n} p_{\bf n} \qquad {\rm with} \qquad
   \dot{\xi}_{\bf n} = {p_{\bf n}^2\over T} - 1\;.     \label{hoover}
\ee
Since our overall accuracy was mainly limited by statistical fluctuations and 
not by integration errors, we used mainly a simple leap-frog integrator 
\cite{evans} with a large step size $dt=0.05$. Test runs with smaller $dt$ 
were made to verify the results (see below). Total integration times 
were typically between $2\times 10^6$ and $2\times 10^8$ units.

\begin{figure}
%Fig 1
\psfig{file=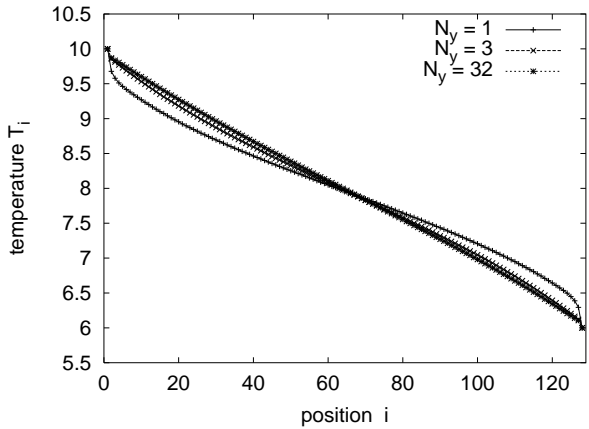,width=5.8cm,angle=270}
\caption{Temperature profiles for lattices with $N_x=128$ and $N_y=1, 3$, and 32.
For these simulations we used a smaller step size $dt = 0.025$.}
\end{figure}

\begin{figure}
%Fig 2
\psfig{file=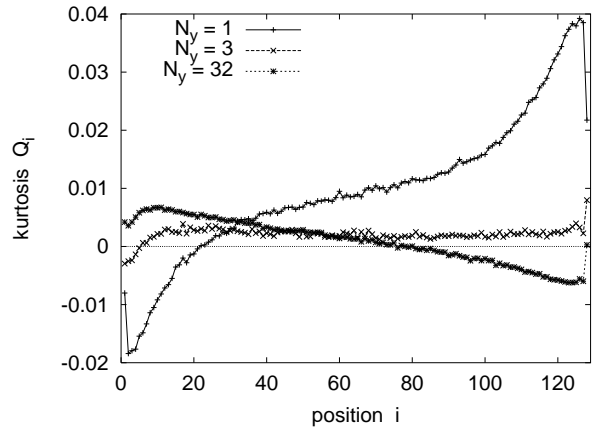,width=5.8cm,angle=270}
\caption{Normalized kurtosis profiles from the same runs used also in Fig.1.}
\end{figure}

Although there exist more sophisticated symplectic integrators 
for Nos{\'e}-Hoover baths \cite{leapfrog}, and although the 
standard form given in Eq.(\ref{hoover}) 
is not symplectic, the straightforward leap-frog was also used for them. 
Apart from simplicity and robustness, our main reason was that LTE
should be strongly violated at the boundaries anyhow, due to the 
large temperature gradients \cite{lepri-review} at the boundaries (see also
Fig.1). If LTE would hold everywhere, and $\kappa$ would depend only on 
$T$ and on the system size, the constancy of the heat flow throughout the 
sample would imply a nearly linear temperature profile (since $\kappa$ depends 
smoothly on $T$). This is obviously not the case. A more direct check of LTE 
is obtained by estimating the normalized kurtosis (fourth cumulant) of the 
momentum distribution, 
\be
   Q_i = {N_y \langle \sum_{j=1}^{N_y} p_{(i,j)}^4\rangle \over
          \left(\langle \sum_{j=1}^{N_y} p_{(i,j)}^2\rangle\right)^2} - 3\;.
\ee
In Fig.2 we show these cumulants for length $N_x=128$ and different widths.
All three curves show that indeed $|Q_i| \ll 1$, verifying approximate LTE 
(for larger $N_x$ we find even smaller $Q_i$, as we should expect from the 
fact that temperature gradients decrease with $N_x$). The detailed curves
depend strongly on $N_y$, but in all cases the kurtosis is largest near 
the boundaries. Thus data obtained from the boundary regions should not 
be used for the analysis anyhow, and integration errors in these regions 
should not matter too much.

Like the kurtosis, local temperatures and heat fluxes were indeed not 
calculated as functions
of the spatial positions $(x, y)$ but for fixed particles: The temperature
of the $i$-th layer is simply $T_i = N_y^{-1} \langle \sum_{j=1}^{N_y} 
p_{(i,j)}^2\rangle$, while the average heat flux through any horizontal 
bond connecting the layers $i$ and $i+1$ is
\be
   J_i = (2N_y)^{-1} \langle \sum_{j=1}^{N_y} (\dot{q}_{(i,j)}
         +\dot{q}_{(i+1,j)} ) F_{(i,j)} \rangle,
\ee
where $F_{(i,j)}$ is the force acting between particles $(i,j)$ and 
$(i+1,j)$. We checked that indeed $J_i$ is independent of $i$, within the 
statistical errors.

\begin{figure}
%Fig 3
\psfig{file=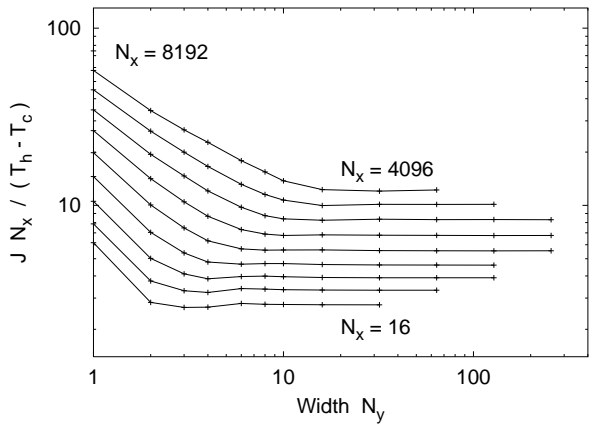,width=5.8cm,angle=270}
\caption{Conductivities defined by Eq.(\ref{global}) for systems with fixed 
   lengths $N_x=32,64,128,\ldots 8192$ against $N_y$ (for $N_x=8192$ there 
   is one single point at $N_y=1$). Statistical and 
   integration errors are less than 2 percent.}
\end{figure}

\begin{figure}
%Fig 4
\psfig{file=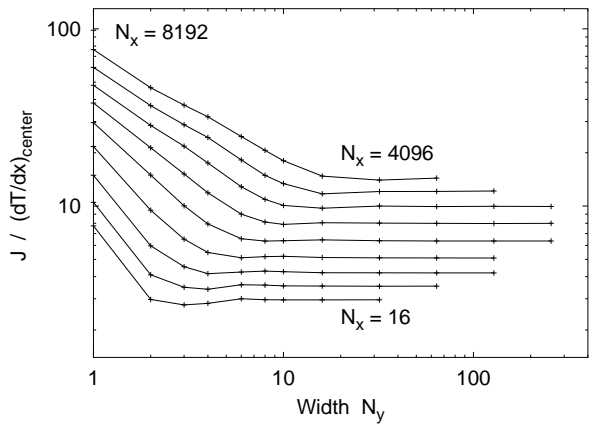,width=5.8cm,angle=270}
\caption{Conductivities defined by Eq.(\ref{local}) from the same runs 
   as the data in Fig.3.}
\end{figure}

\begin{figure}
%Fig 5
\psfig{file=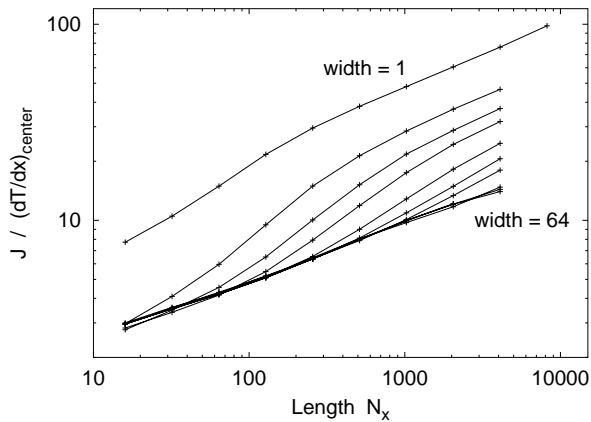,width=5.8cm,angle=270}
\caption{Same data as in Fig.~4, but plotted against $N_x$ and with data 
   for the same $N_y$ connected by lines.}
\end{figure}

Conductivities obtained by dividing the flux by the globally averaged 
temperature gradient,
\be
   \kappa_{\rm global} = {\sum_{i=1}^{N_x} J_i \over (T_h - T_c) }\;,  
                         \label{global}
\ee
are shown in Fig.~3, while conductivities obtained by using the gradient 
averaged only over the central half of the lattice, 
\be
   \kappa_{\rm center} = {\sum_{i=1}^{N_x} J_i \over N_x (dT/dx)_{\rm center}}
                         \label{local}
\ee
with $(dT/dx)_{\rm center} = 2 (T_{3N_x/4} - T_{N_x/4})/N_x$, are shown in
Figs.~4 and 5. There are small but significant differences 
between the two definitions of $\kappa$. In general, the plots for 
$\kappa_{\rm center}$ show slightly more structures. We argue that they are 
more relevant, since they are less affected by boundary effects. We show the 
data for $\kappa_{\rm global}$ also, because they are much less noisy.

From Figs.~3 and 4 we see that $\kappa$ saturates for large $N_y$. The 
saturation values increase roughly power-like with $N_x$. For intermediate
$N_y$ there are very shallow minima, at $(N_y)_{\rm min} \sim N_x^{0.35}$. 
A rough fit to the data for $N_y > 2$ is obtained with a scaling ansatz
\be
    \kappa = N_x^{\alpha'} \phi(N_y/N_x^\beta) \;,
            \qquad \alpha' = 0.26,\;\beta = 0.35\;.          \label{scaling}
\ee
with $\phi(x) \approx const$ for $x\gg 1$ and $\phi(x) \sim x^{(\alpha'-\alpha)/\beta}$
for $x\to 0$. Thus the cross-over happens at $r^* \sim N_x^{1-\beta} = N_x^{0.65}$.
As seen from Fig. 5, both the 1-d behaviour ($r=\infty$)
and the behaviour at $r \ll r^*$ are rougly power like, but with different
powers. In between, $\kappa$ increases even faster with $N_x$ (for fixed 
$N_y$). 

For fixed aspect ratios this leads to a decrease of $\kappa$ with
$N_x$ for small $N_x$ ($r > r^*(N_x)$), and to a power increase for large 
$N_x$ (see Fig.~6). Indeed one sees definite deviations from a pure power 
law, even for large $N_x$. This means that $\kappa$ increases asymptotically
less fast than suggested by a simple least square fit to the present data 
(which would give the exponent $\alpha'=0.26$ used in Eq.(\ref{scaling})). 
Thus the true asymptotic exponent $\alpha(2{\rm d})$ for 2-d lattices is 
less than $\alpha'$.  Any detailed extrapolation is of course uncertain, 
but our best estimate is $\alpha(2{\rm d}) = 0.22\pm 0.03$. 

\begin{figure}
%Fig 6
\psfig{file=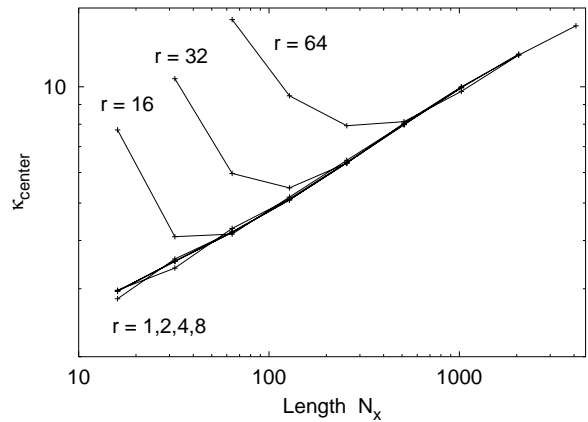,width=5.8cm,angle=270}
\caption{Conductivities defined by Eq.(\ref{local}) for fixed aspect ratios.}
\end{figure}

\begin{figure}
%Fig 7
\psfig{file=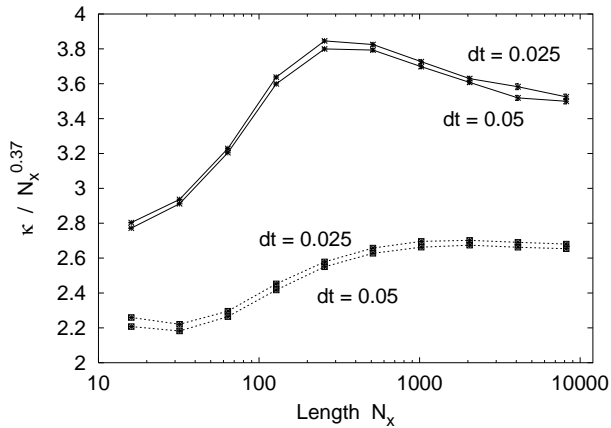,width=5.8cm,angle=270}
\caption{The data for $N_y=1$, divided by $N_x^{0.37}$. The upper two curves 
   are for $\kappa_{\rm center}$, the lower for $\kappa_{\rm global}$.}
\end{figure}

In order to see more precisely the behaviour of 1-d chains (where we have 
much higher statistics than for $N_y>1$), and to see more clearly the errors
made by the finite step size $dt$, we show $\kappa / N_x^{0.37}$ in Fig.~7.
Indeed the uncertainty in the definition of $\kappa$ is much 
more important than the integration error. Although the latter is much 
larger than the statistical error (since we obtained very high statistics
for $N_y=1$), it has hardly any effect on the scaling of $\kappa$. While 
the curves for $\kappa_{\rm global}$ seem to flatten for large $N_x$, those for 
$\kappa_{\rm center}$ seem to turn upward again for very large $N_x$, after
having bent downward for intermediate $N_x$. Essentially the same behaviour 
was seen for the 1-d gas with alternating masses in \cite{grass-lei}. 
Again any extrapolation is rather uncertain, our best estimate is
$\alpha(1{\rm d}) = 0.37\pm 0.01$.

In summary, we have studied the heat conduction on rectangular 2-d FPU 
lattices with periodic lateral boundary conditions, including lattices 
with zero width (i.e. linear chains). We argued that the divergence of 
the conductivity with system
size should be universal, so that our results should apply also, among 
others, to carbon nanotubes. One striking result is that we do not see
the logarithmic increase of $\kappa$ found in \cite{lippi}. At least for 
system sizes up to five times the size of those studied in 
\cite{lippi}, the increase is power-like, with an exponent larger than 0.2.
This increase flattens somewhat for even larger lattices, but our best
asymptotic estimate is still around 0.2. For $d=1$ our estimate of the 
exponent $\alpha$ is slightly smaller than previous ones, but 
compatible with them.

Our other striking result is that the cross-over from 1-d to 2-d behaviour 
happens at an aspect ratio which diverges with system size. Thus it might
be difficult to see the true 1-d behaviour in carbon nanotubes. But this 
does not mean that $\kappa$ should not increase with their length, quite
to the contrary: The increase at presently achievable tube lengths might 
even be faster than both the asymptotic 1-d and 2-d power laws, in view
of the special shape of the cross-over. Since $\kappa$ is decreasing with
the width $L_\perp$ for large aspect ratios, two tubes with small radius
conduct better than one tube with twice the radius. Thus, if one wants to 
maximize the heat flux for a given amount of carbon and a given length of 
the tubes, one should make the tubes as narrow as possible. 

Finally we would like to point out that the effect we discussed in this 
paper is closely related to the Casimir force. The latter deals with the 
change in energy of a cavity due to elimination of vacuum modes by the 
cavity walls. The enhanced conductivity of narrow strips compared to 
the bulk is due to the elimination of transverse phonons (i.e. phonons with
$p_y \ne 0$), the interactions with which otherwise would limit the free 
path length of soft longitudinal phonons which contribute to the energy 
transport.

We are indebted to Roberto Livi and Antonio Politi for very helpful 
correspondence, and to Walter Nadler for numerous discussions.

\end{document}